\documentclass{ws-ijmpc}

\def\Z{{\mathbb Z}}

\begin{document}

\markboth{Murgu\'ia, Mej\'ia Carlos, Rosu $\&$ Flores-Era\~na}
{Improvement and Analysis of a Pseudo Random Bit Generator by Means of
        Cellular Automata}

\catchline{}{}{}{}{}

 \title{\bf IMPROVEMENT AND ANALYSIS OF A PSEUDO RANDOM BIT GENERATOR BY MEANS OF
        CELLULAR AUTOMATA}

   \author{J. S. MURGU\'IA \footnote{Corresponding author}}
   \address{\it UASLP, Universidad Aut\'onoma de San Luis Potos\'i\\
          \'Alvaro Obreg\'on 64, 78000 San Luis Potos\'i, S.L.P., M\'exico\\
          IPICyT, Instituto Potosino de Investigacion Cientifica y Tecnologica\\
          Apartado Postal 3-74 Tangamanga, 78231 San Luis Potos\'{\i}, M\'exico\\
          ondeleto@uaslp.mx}

  \author{M. MEJ\'IA CARLOS}
  \address{\it Instituto en Investigaci\'on en Comunicaci\'on \'Optica, UASLP\\
       \'Alvaro Obreg\'on 64, 78000 San Luis Potos\'i, S.L.P., M\'exico\\
       mmejia@cactus.iico.uaslp.mx}

  \author{H.C. ROSU}
  \address{IPICyT,
  Instituto Potosino de Investigacion Cientifica y Tecnologica\\
  Apartado Postal 3-74 Tangamanga, 78231 San Luis Potos\'{\i},
  M\'exico\\
  hcr@ipicyt.edu.mx}

  \author{G. FLORES-ERA\~NA}
  \address{\it Instituto en Investigaci\'on en Comunicaci\'on \'Optica, UASLP\\
  \'Alvaro Obreg\'on 64, 78000 San Luis Potos\'i, S.L.P., M\'exico\\
   gustavo.flores.erana@gmail.com}

\maketitle

  \begin{history}
    \received{Day Month Year}
    \revised{Day Month Year}
  \end{history}

    \begin{abstract}

    In this paper, we implement a revised pseudo random bit generator based on a rule-90 cellular automaton. For this purpose, we introduce a sequence matrix $H_N$ with the aim of calculating the pseudo random sequences of $N$ bits employing the algorithm related to the automaton backward evolution. In addition, a multifractal structure of the matrix $H_N$ is revealed and quantified according to the multifractal formalism.
     The latter analysis could help to disentangle what kind of automaton rule is used in the randomization process and therefore it could be useful in cryptanalysis. Moreover, the conditions are found under which this pseudo random generator passes all the statistical tests provided by the
    National Institute of Standards and Technology (NIST).

  \keywords{Cellular automata, pseudo-random generator; multifractal spectrum.}
 \end{abstract}

\ccode{PACS numbers.: 05.40.-a., 05.45.-a., 05.45.Tp}

\section{Introduction}

  Random numbers constitute one of the main
  ingredients in a great number of applications such as
  cryptography, simulation, games, sampling, and so on.
  Cellular automata (CA) offer a number
  of advantages over other methods as random number
  generators (RNG), such as algorithmic simplicity and
  easy hardware implementation.
  In fact, CA are highly parallel and distributed systems
  which are able to perform complex computations.

  Over the last years, researchers have applied cellular automata
  in pseudo-random number generation (PRNG)
  \cite{Urias,Marcela-DCDS,Marcela-PhD,sipper,Seredynski}.
   These PRNGs must possess a number of
   properties if they are to be used
   for cryptographic application.
   The most important from this point of view are good results
   on standard statistical tests of randomness, computational
   efficiency, a long period, and reproducibility of the
   sequence~\cite{zied}.
   In many encryption systems, PRNGs
   are used to get keys, which are generated from an initial seed,
   and they are reproducible if the same seed is used.

   Wolfram was the first to apply the one-dimensional elementary cellular automata (ECA) to obtain PRNGs \cite{wolfram}.
   He considered only the use of the rule 30 in one dimension with radius 1. Other authors have used non-uniform ECA \cite{Seredynski,Hortensius,Nandi}, where they have found that
   the quality of the latter PRNGs was better than the quality of Wolfram's system.

   Despite these works demonstrate an improvement in the quality of PRNGs, this study is devoted to an extension of
   the analysis of the evolutionary technique for getting PRNGs based on a uniform ECA
   with rule 90~\cite{Marcela-DCDS}. Namely, we consider a modification of the generator producing PRGNs.
   The initial proposal has been never implemented and studied in terms of the
   sequence matrix $H_N$, which is used here to generate recursively the pseudo random sequences. The time series of the row sums of the matrix $H_N$ are also analyzed within the multifractal formalism because they could be a possible useful feature for cryptanalysis of these types of PRNGs.
   In order to check the quality of this ECA-PRNG, the generated sequences are evaluated statistically
   by the NIST suite. The generated sequences, in length terms, pass all the statistical tests proposed by NIST.
   Our results suggest that this generator in its two versions that we discuss in the following fit naturally in the present digital communication
   systems and achieve high levels of performance.

\section{Elementary Cellular Automata}
    \label{s-CA}

  The ECA can be considered as
  discrete dynamical systems that evolve in discrete time steps.
  The state space of a CA of size $N$ is the set
  $\Omega=\Z^N_k$ of all sequences of $N$ cells that take
  values from $\Z_k=\{0,1,\ldots,k-1\}$, where its evolution
  is defined by the repeated iteration of an
  evolution operator $\mathcal{A}:\Z^N_k \to \Z^N_k$.
  In this paper, we consider $k=2$ where $\Z$ is the
  set of integers.
  An automaton state $\underline{x} \in \Z^\Z_k$ has coordinates
  $(\underline{x})_i = x_i \in \Z_2$ with $i\in \Z$, and
  the automaton state at time $t\geq 0$ is denoted
  by $\underline{x}^t \in \Z^\Z_2$ and its evolution
  is defined iteratively by the rule $\underline{x}^{t+1} = A (\underline{x}^t)$.
  Starting from the initial state $\underline{x}^0$, the automaton
  generates the forward space-time pattern $\mathbf{x} \in \Z^{\Z \times \mathbb{N}}_2$
  with state $(\mathbf{x})^t = \underline{x}^t = A^t(\underline{x}^0)$
  reached at from $\underline{x}^0$ after $t\in \mathbb{N}$ time steps.
  $\mathbb{N}$ denotes the set of nonnegative integers.

  One can see that the time, space, and states of
  this system take only discrete values.
  The ECA considered evolves according to the local rule
  $x_i^{t+1} = \mathcal{A}_L (x_{i-1}^t, x_{i}^t, x_{i+1}^t) = [x_{i-1}^t + x_{i}^t] \rm{mod}\ 2$,
  which corresponds to the rule 90.
  The following is the lookup table of rule 90.

   \medskip

     \begin{center}
     \begin{tabular}{lllllllll}
         \hline
         \multicolumn{1}{|l|}{Number} &
         \multicolumn{1}{|c|}{7} &\multicolumn{1}{|c|}{6} & \multicolumn{1}{|c|}{5} & \multicolumn{1}{|c|}{4} &
         \multicolumn{1}{|c|}{3} & \multicolumn{1}{|c|}{2} & \multicolumn{1}{|c|}{1} & \multicolumn{1}{|c|}{0} \\
         \hline
         \multicolumn{1}{|l|}{Neighborhood} &
         \multicolumn{1}{|c|}{111} & \multicolumn{1}{c|}{110} & \multicolumn{1}{c|}{101} & \multicolumn{1}{c|}{100} & \multicolumn{1}{c|}{011} & \multicolumn{1}{c|}{010} & \multicolumn{1}{c|}{001} & \multicolumn{1}{c|}{000} \\
         \hline
         \multicolumn{1}{|l|}{Rule result} &
         \multicolumn{1}{|c|}{0} & \multicolumn{1}{c|}{1} & \multicolumn{1}{c|}{0} & \multicolumn{1}{c|}{1} & \multicolumn{1}{c|}{1} & \multicolumn{1}{c|}{0} & \multicolumn{1}{c|}{1} & \multicolumn{1}{c|}{0} \\
         \hline
     \end{tabular}
     \end{center}
  \medskip

     The third row shows the future state of the cell
     if itself and its neighbors are in the arrangement shown
     above in the second row. In fact, a
     rule is numbered by the unsigned decimal equivalent of the binary expression in
     the third row.
     When the same rule is applied to update cells of ECA,
     such ECA are called uniform ECA; otherwise the
     ECA are called non-uniform or hybrids.
   It is important to observe that the evolution rules of ECA
  are determined by two main factors, the rule and the initial
  conditions.

  \section{Pseudo Random Sequences Generator}
      \label{s-PRNG}

\begin{figure}[ph]
      \centerline{\psfig{file=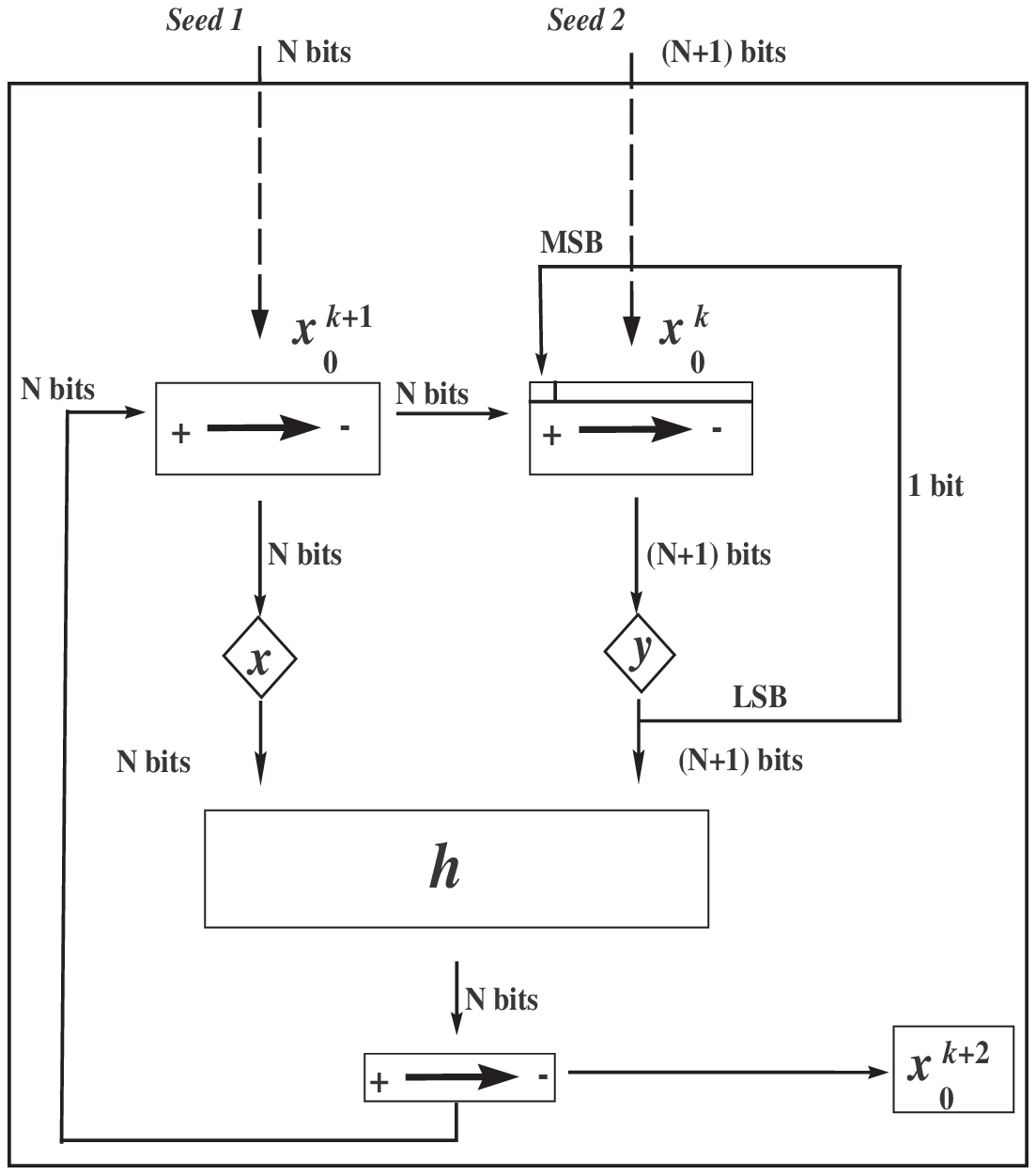,width=7cm, height=8cm}}
      \vspace*{8pt}
      \caption{
         Basic form of the pseudo-random number generator. MSB and LSB
         correspond to the most significant bit and the least significant bit, respectively.}
        \label{fig-KeysG}
      \end{figure}

   In Ref.~\refcite{Marcela-DCDS}, Mej\'ia and Ur\'ias
   presented an ergodic and mixing transformation
   of binary sequences in terms of a cellular automaton, which
   is the main element of a pseudo-random generator number (PRGN).
   To implement numerically the PRGN in its basic form,
   we follow their algorithm,
   which is shown in
   Fig. \ref{fig-KeysG}. At first, the key generator requires two seeds,
   $\mathbf{x}=\mathbf{x}_0^{k+1}$, of $N$ bits, and
   $\mathbf{y}=\mathbf{x}_0^{k}$,
   of $(N+1)$ bits, which are the input of function
   $\mathbf{t} = h(\mathbf{x},\mathbf{y})$.
   The seeds are
   $\mathbf{x} = \{x_1,x_2, x_3, \ldots, x_N\}$ and
   $\mathbf{y} = \{y_1,y_2, y_3, \ldots, y_{N+1}\}$,
   and the first number generated of $N$ bits is the sequence output of function $h$,
   $\mathbf{t} = x_0^1 = \{t_1,t_2, t_3, \ldots, t_N\}$.
   Now this sequence is feeding back to the input,
   which becomes
   the next value of
   $\mathbf{x}$, and the previous
   value of $\mathbf{x}$
   becomes the initial bits of the new $\mathbf{y}$,
   where the missing bit is the least significant bit (LSB) of the previous $\mathbf{y}$,
   which becomes the most significant bit (MSB) of this sequence,
   and the same procedure is iterated repeatedly.

\begin{figure}[ph]
      \centerline{\psfig{file=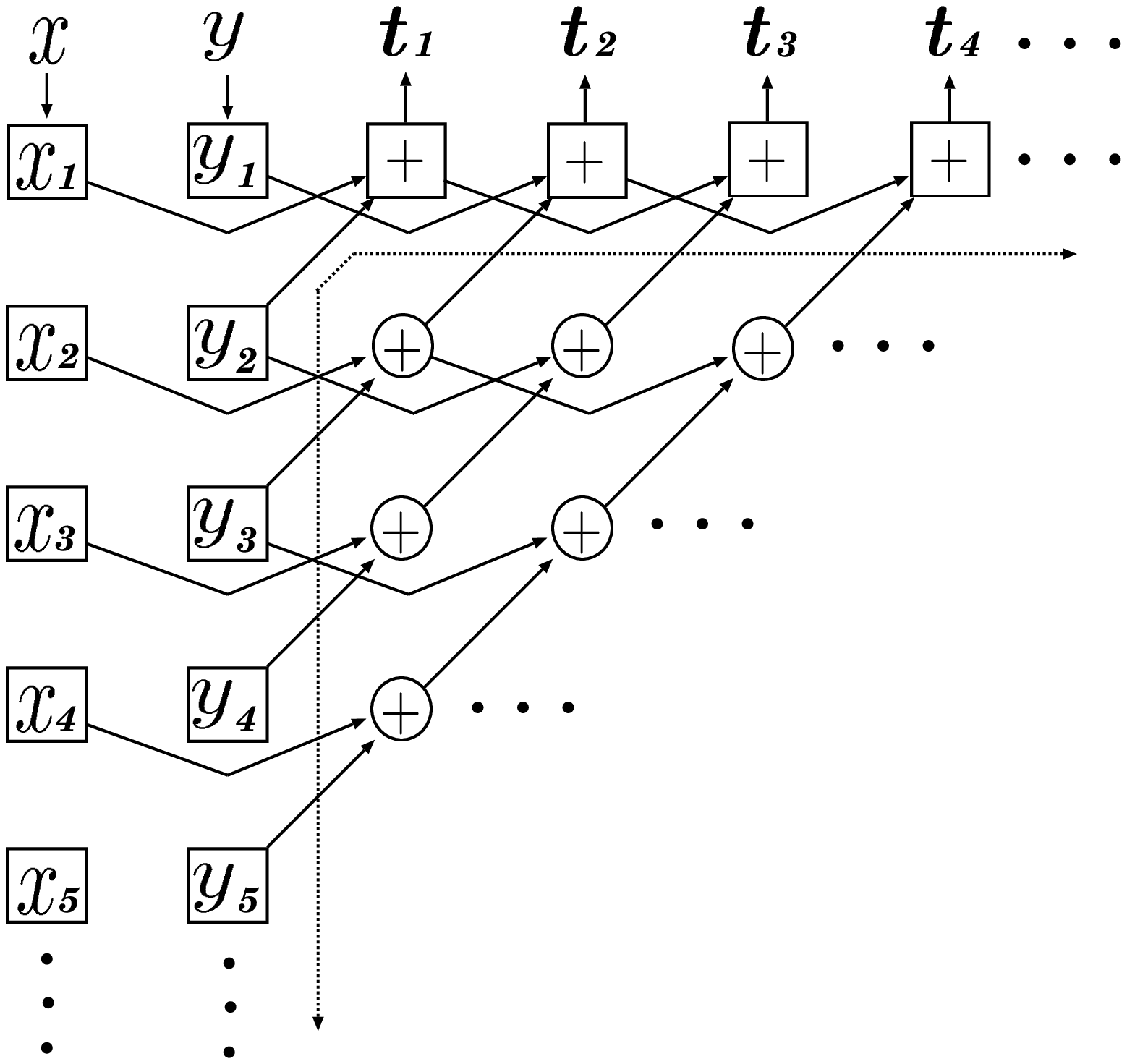,width=5cm, height=4.5cm}}
      \vspace*{8pt}
      \caption{
         Generation of a pseudo-random key with
         input $(\mathbf{x},\mathbf{y})$ and
         output $\mathbf{t} = h(\mathbf{x},\mathbf{y})$.
         }
        \label{fig-Tgeneration}
      \end{figure}

\begin{figure}[ph]
      \centerline{\psfig{file=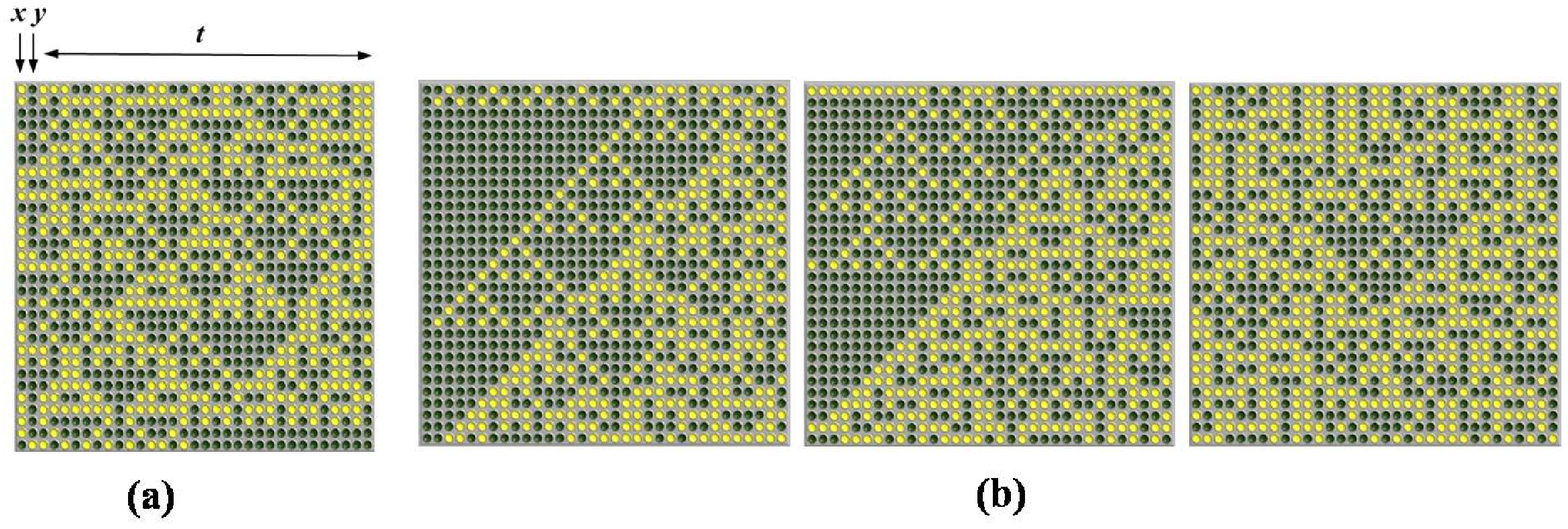,width=13.5cm, height=4.5cm}}
      \vspace*{8pt}
      \caption{
          Complete backward evolution of the ECA to generate a
          random sequence of 31 bits with (a) one transformation according to Eq.~(\ref{eq-LlavesG}),
           and with (b) three transformations according to the modified generator given in Eq.~(\ref{eq-matriz-Generador3}). We display the 
           sequences $p_k$, $q_k$, and $x_k$.
          }
         \label{fig-Evolution10}
       \end{figure}

   The previous description to compute
   the function $\mathbf{t}= h(\mathbf{x},\mathbf{y})$
   requires
   that the cellular automaton runs backwards in time as is depicted
   in Fig. \ref{fig-Tgeneration}. The symbol
   of a circled $+$
   represents a XOR gate and the connectivity
   of gates follows the automaton rule.
%
   However, this way to compute the pseudo-random sequences
   is not efficient since it requires the application of the
   local rule of the automaton at all points
   in a lattice of the order of $N^2$, where $N$ is the
   number of bits considered in the generation process.
   Fig.~\ref{fig-Evolution10} (a) shows a complete
   evolution in the lattice of the generator considered
   with $N=31$ bits, where
   the two left columns comprise
   the seed $(\mathbf{x},\mathbf{y})$,
   the top row is the resulting pseudo random key sequence $\mathbf{t}$, and
   the intermediate elementary computations are the rest.
   A clear dot represents a bit value of 1, whereas a dark dot
   corresponds to a bit value of 0.

%
   To overcome this, Mej\'ia and Ur\'ias~\cite{Marcela-DCDS} formulated an efficient algorithm
   that gets rid of the intermediate variables and produces
   boolean expressions for the coordinates of the output
   sequence $\mathbf{t}= h(\mathbf{x},\mathbf{y})$ in terms of the input $(\mathbf{x},\mathbf{y})$.
   This algorithm
   offers a boolean representation of $h$, without
   intermediate steps, in terms of some ``triangles'' in the underlying
   lattice. The pattern of triangles is observed in
   Fig.~\ref{fig-Evolution10}(a). In fact, it is well known that the evolution of rule 90 has
   the appearance of a Sierpinski triangle when responding to an impulse, i.e., when the first row is all 0s with
   a 1 in the center.

   We introduce now the sequence matrix $H_N$, which computes the pseudo-random sequences of $N$ bits.
   This matrix has dimensions of $ (2N+1)\times(2N+1)$
   and is formed by the matrices $H_{N_t}$ and $H_{N_b}$, which
   constitute the top and bottom parts
   of $H_N$, that is, $H_N = \big( H_{N_t}; H_{N_b} \big)$.
   The matrix $H_{N_t}$ has dimensions of
   $ N \times(2N+1)$ elements and it is generated
   initially from vectors
   $\mathbf{v} =[v_1, \, 0, \, \ldots, \, 0, \, v_{N+2}, \, \ldots, \, 0]$
   and
   $\mathbf{w} =[0, \, w_2, \, 0, \, \ldots, \, w_{N+1}, \, 0, \, w_{N+3}, \, \ldots, \, 0]$,
   where the components $v_1, v_{N+2}, w_2, w_{N+1}$ and $w_{N+3}$ have
   a value of 1,
   and $N$ is the number of bits, i. e., $\mathbf{v}$ and
   $\mathbf{w}$ are vectors with $(2N+1)$ elements.
   The vectors $\mathbf{v}$ and $\mathbf{w}$ constitute
   the two first rows of the matrix $H_{N_t}$ and the $(N-2)$ rows
   are generated applying an addition modulo 2 operation
   of the two previous rows, with the elements of the previous
   row
   shifted to the right by one position. The
   matrix $H_{N_t}$, of dimensions $(N+1)\times(2N+1)$,
   has a simpler form, an identity matrix
   in the first $(N+1)$ columns and zeros in the rest.
   For instance, for $N=3$ we have that the top and bottom
   matrices are

   \begin{equation}
        \label{ec-matriz-Hs3}
        H_{3_t}=\left(
            \begin{array}{ccccccc}
                 1 & 0 & 0 & 0 & 1 & 0 & 0 \\
                 0 & 1 & 0 & 1 & 0 & 1 & 0 \\
                 1 & 0 & 1 & 0 & 0 & 0 & 1
            \end{array}%
                \right),
            \qquad
        H_{3_b}=\left(
            \begin{array}{ccccccc}
            1 & 0 & 0 & 0 & 0 & 0 & 0 \\
            0 & 1 & 0 & 0 & 0 & 0 & 0 \\
            0 & 0 & 1 & 0 & 0 & 0 & 0 \\
            0 & 0 & 0 & 1 & 0 & 0 & 0
            \end{array}%
            \right),
    \end{equation}

  then the matrix $H_3$ is

     \begin{equation}
        \label{eq-matriz-H3}
        H_3 =
            \left(
                \begin{array}{c}
                  H_{3_t}  \\
                  H_{3_b}
                \end{array}%
                \right)
                =\left(
                    \begin{array}{ccccccc}
                    1 & 0 & 0 & 0 & 1 & 0 & 0 \\
                    0 & 1 & 0 & 1 & 0 & 1 & 0 \\
                    1 & 0 & 1 & 0 & 0 & 0 & 1 \\
                    1 & 0 & 0 & 0 & 0 & 0 & 0 \\
                    0 & 1 & 0 & 0 & 0 & 0 & 0 \\
                    0 & 0 & 1 & 0 & 0 & 0 & 0 \\
                    0 & 0 & 0 & 1 & 0 & 0 & 0
            \end{array}%
            \right).
    \end{equation}

   Notice that $H_{N_t}$ computes the pseudo random key sequence,
   whereas $H_{N_b}$ the feedback sequence. Therefore,
   once selected the number $N$ of bits of sequences, we can
   generate the pseudo random sequences of $N$ bits
   with the help of the matrix $H_N$

   \begin{equation}\label{eq-LlavesG}
        \mathbf{U}_{k+1} = H_N \mathbf{U}_k, \quad k=1,2,\ldots
   \end{equation}

   where $\mathbf{U}_k= [\mathbf{x} \ \mathbf{y}]^T$ corresponds to the
   first inputs of function $h$, and
   $\mathbf{U}_{k+1}$ is composed by the next inputs of $h$;
   note that $\mathbf{U}_{k+1}$ is formed by the
   generated pseudo random key and the feedback sequence.

   \subsection{Modified generator}

\begin{figure}[ph]
      \centerline{\psfig{file=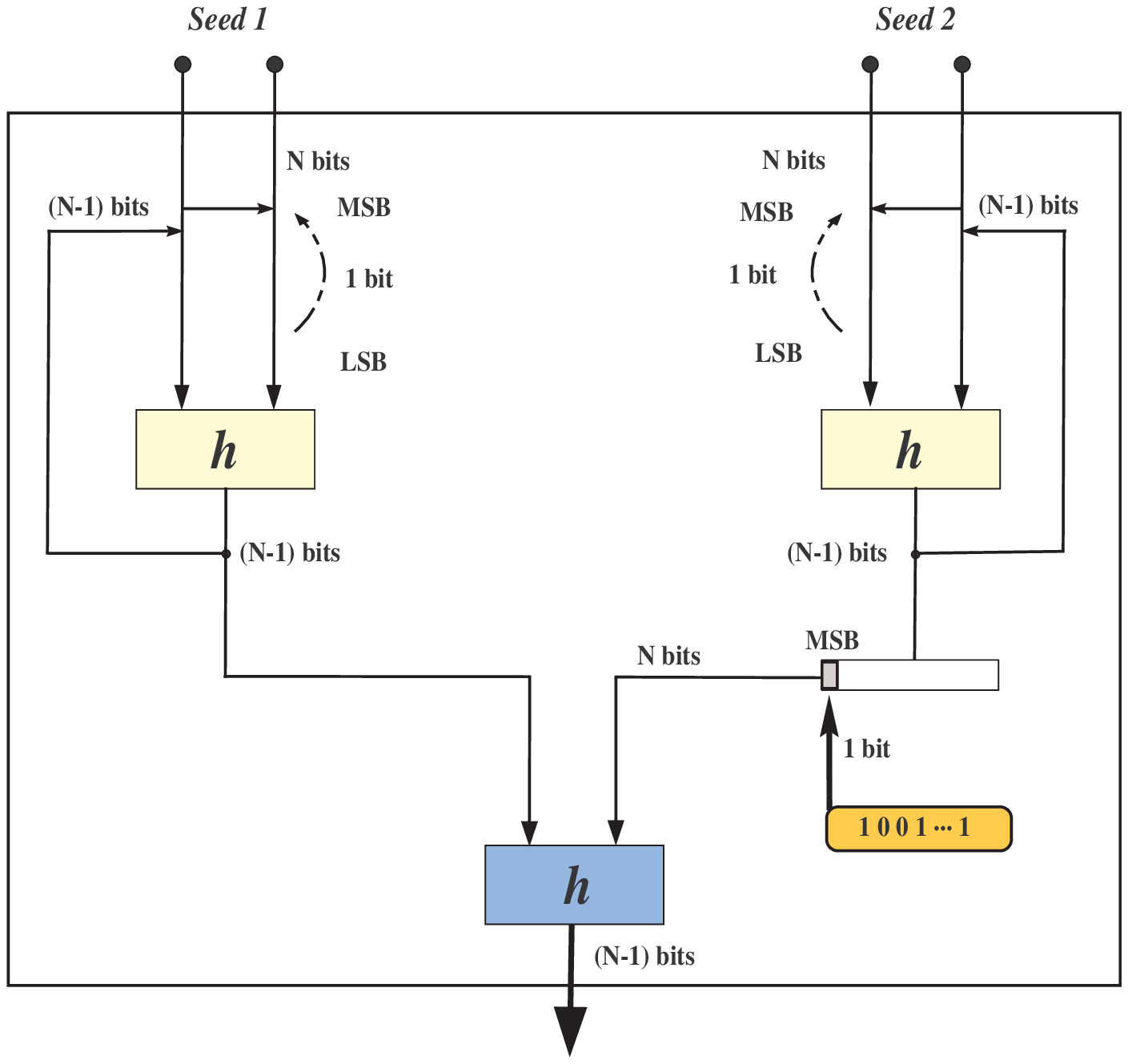,width=8cm, height=8cm}}
      \vspace*{8pt}
      \caption{
         A generating scheme consisting of three coupled transformations.}
        \label{fig-Keys3F}
      \end{figure}

    As was pointed out in Ref.~\refcite{Marcela-DCDS},
   a generating
   scheme consisting of three coupled transformations $h$
   is proposed to attain an asymptotically unpredictable
   generator under a random search attack.
   This proposal
   is shown in Fig.~\ref{fig-Keys3F}, and it is explained briefly. %
  Inside the new generator two copies of the
  basic transformation $h$ are iterated autonomously
  from their initial words generating two sequences,
  $\{p^k\}_{k\geq0}$ and $\{q^k\}_{k\geq0}$.
  The third copy, called the $x$-map, is
  iterated in a slightly different manner, the function
  $h$ in the $x$-map is driven by the autonomous
  $p$-map and $q$-map according to
  $x^k = h(p^k,q^k)$. The three maps generate pseudo random
  sequences, but just the $x$ sequence is released.
  In order to prevent predictability, the first two words
  are generated, used and destroyed inside this
  key generator, therefore they are not available
  externally. Since the sequences $p^k$ and $q^k$
  have a length of $N$ bits, and
  the required inputs of the $h$ transformation
  must be one of $N$ bits and other of $(N+1)$ bits,
  the missing bit is obtained applying
  an addition modulo 2 operation between the two respective
  LSB's that become the MSB's of their respective
  previous inputs of maps.
  Of course, there exists different manners to
  generate this missing bit, but we consider
  this way.
  The above scheme has been just proposed, but it has
  not been implemented and studied in terms of the matrix sequence.
  The new pseudo-random keys are computed
  as

   \begin{equation}
        \label{eq-matriz-Generador3}
        \mathbf{X}_N = H_{N_t} \mathbf{V}_N
    \end{equation}

  where $\mathbf{X}_N = \{x_1, \, x_2, \, \ldots, \, x_N\}^T$,
  $H_{N_t}$ is the top matrix of $H_N$, and
  $\mathbf{V}_N = \{p_1, \, \ldots, \, p_N, \, q_1, \, \ldots, \, q_{N+1} \}^T$.
  For example, considering $N=3$, we have

     \begin{equation}
        \label{eq-matriz-Generador3-n3}
        \mathbf{X}_3 =
            \left(
                \begin{array}{c}
                  x_1  \\
                  x_2  \\
                  x_3
                \end{array}%
                \right)
                =\left(
                    \begin{array}{ccccccc}
                    1 & 0 & 0 & 0 & 1 & 0 & 0 \\
                    0 & 1 & 0 & 1 & 0 & 1 & 0 \\
                    1 & 0 & 1 & 0 & 0 & 0 & 1
            \end{array}%
            \right)
            \left(
                \begin{array}{c}
                  p_1  \\
                  p_2  \\
                  p_3  \\
                  q_1  \\
                  q_2  \\
                  q_3  \\
                  q_4
                \end{array}%
                \right)
                = H_{3_t}\mathbf{V}_3,
    \end{equation}

  where we calculate $p_i$ and $q_i$ as it was explained above.
  Fig.~\ref{fig-Evolution10} (b) shows the complete evolution of this
  modified generator for 31 bits, with the $p$-map on the left side, the $q$-map on the
  middle, and the $x$-map on the right side.

   \subsection{Multifractal properties of the matrix $H_N$}

   Since the evolution of the sequence matrix $H_N$
   is based on the evolution of the CA rule 90, the structure of the patterns of bits of the latter are directly reflected in the structure of the entries of $H_N$. There is recent literature on the multifractal properties of cellular automata for some set of rules, see \cite{sanchez,sanchez2,alemanes}.
   In Ref.~\refcite{murguia}, we used the technique of detrended fluctuation analysis
   based on the discrete wavelet transform (WMF-DFA) to quantify the intrinsic multifractal behavior of the ECAs for rules 90, 105, and 150. 

   Here, in the same spirit as in Ref.~\refcite{murguia}, we analyze the sum of ones in the sequences of the rows of  the matrix $H_N$ with
   the db-4 wavelet, a wavelet function that belongs to the Daubechies family~\cite{Mallat}.

      \begin{figure}[ph]
      \centerline{\psfig{file=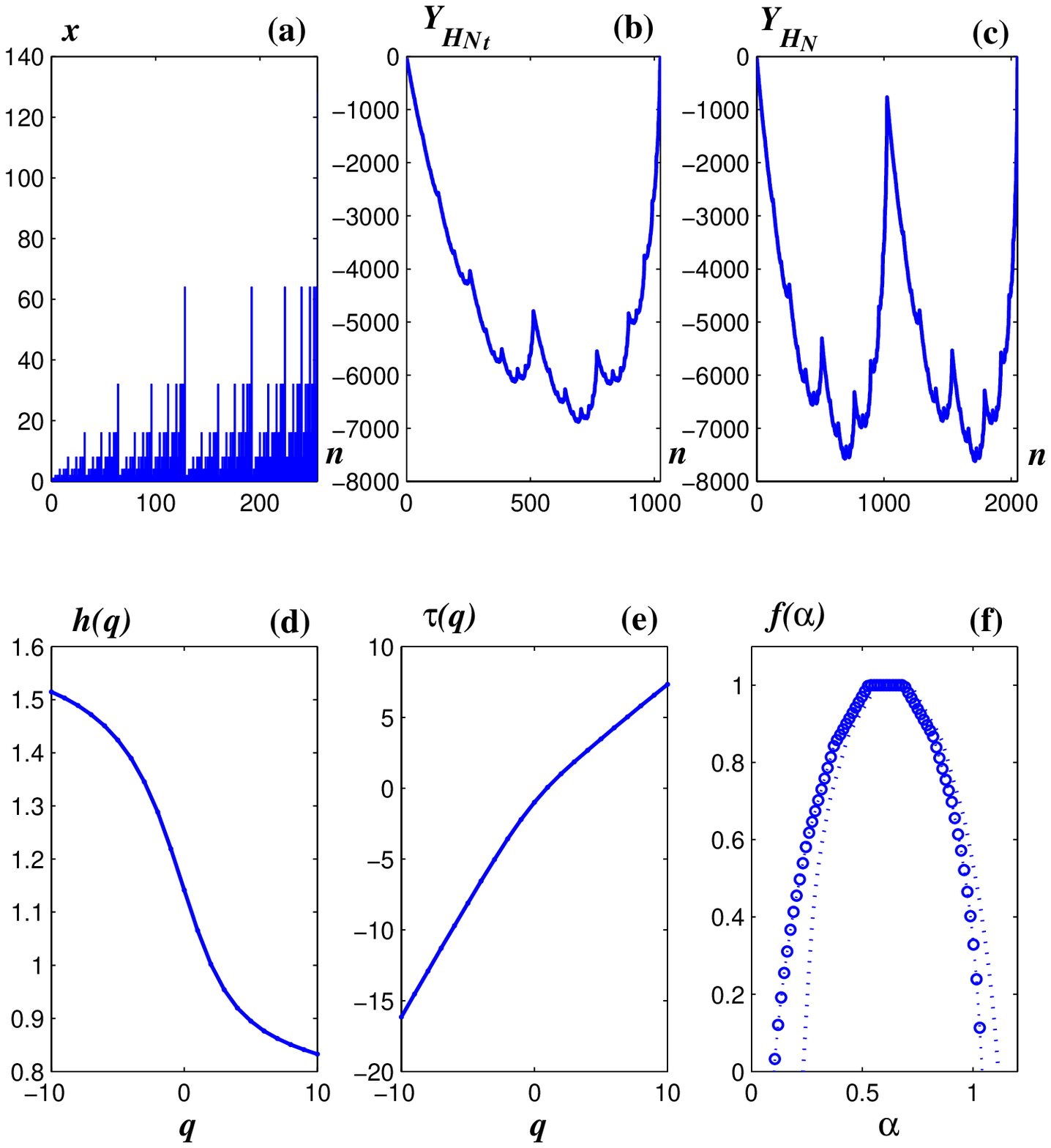,width=10cm, height=12cm}}
      \vspace*{8pt}
      \caption{
          (a) Time series of the row signal of $H_{1023}$. Only the first $2^8$ points
           are shown of the whole set of $2^{10}-1$ data points.
           Profiles of the row signal of (b)$H_{N_t}$ and (c) $H_{N}$. (d) Generalized Hurst exponent $h(q)$.
           (e) The $\tau$ exponent, $\tau(q)=qh(q)-1$.
           (f) The singularity spectrum $f(\alpha)=q\frac{d\tau(q)}{dq}-\tau(q)$.
           The calculations of the multifractal quantities $h$, $\tau$, and $f(\alpha)$
           are performed with the wavelet-based WMF-DFA. Dotted points
           correspond to the row signal of $H_N$.}
         \label{fig-MF10}
       \end{figure}

\begin{figure}[ph]
      \centerline{\psfig{file=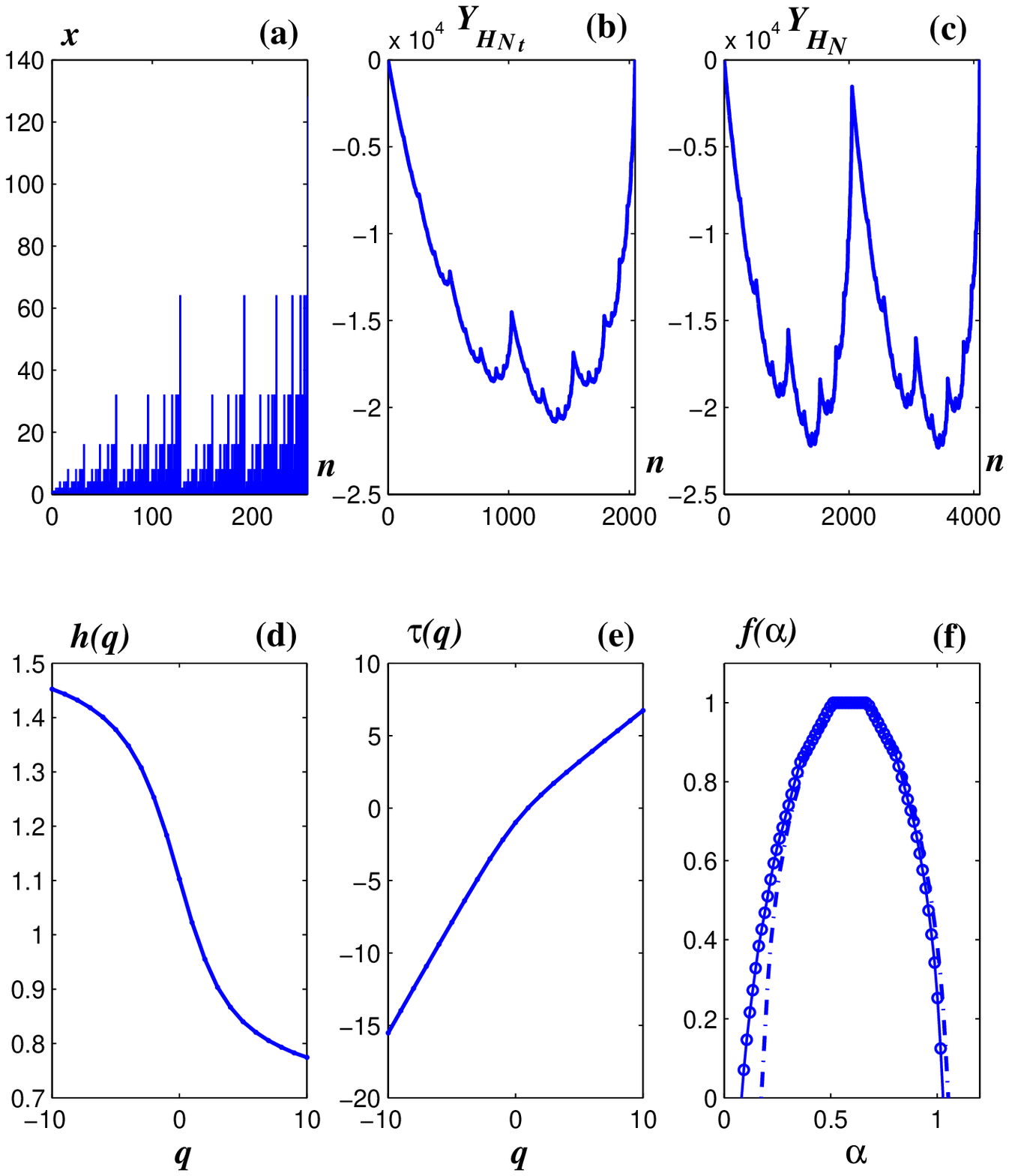,width=10cm, height=10cm}}
      \vspace*{8pt}
      \caption{
        Same plots as in Fig.~\ref{fig-MF10} but for
        $H_{2047}$.}
         \label{fig-MF11}
       \end{figure}

  The results for two row sums, $H_{1023}$ and $H_{2047}$,
  are illustrated in Figs.
  \ref{fig-MF10}-\ref{fig-MF11}. We confirm the multifractality
  of both time series since we get a $\tau$ spectrum
  with two slopes in both cases.
  The strength of the multifractality is roughly
  measured with the width
  $\Delta \alpha = \alpha _{\rm max}- \alpha_{\rm min}$
  of the parabolic singularity spectrum $f(\alpha)$ on the $\alpha$ axis.
  For the case of $H_{1023}$ the width
  $\Delta \alpha_{H_{1023}} = 1.16 - 0.212 =  0.948$,
  and the most ``frequent'' singularity  occurs
  at $\alpha_{\rm{mf}_{H_{1023}}} = 0.694$,
  whereas for $H_{2047}$,
  $\Delta \alpha_{H_{2047}} = 1.12 - 0.145 = 0.975$,
  and $\alpha_{\rm{mf}_{H_{2047}}} = 0.638$. We notice
  that both the strongest singularity,
  $\alpha_{\rm{min}}$, and the weakest singularity,
  $\alpha_{\rm{max}}$, are very similar as well as
  the most ``frequent'' singularity. These results
  are in a good agreement with those obtained
  in Ref.~\refcite{murguia} for the rule 90, although
  the spectra of the top matrix present a
  slight shifting to the right.
  In fact, this behavior is more evident
  in the row signals of $H_N$ (Figs.~\ref{fig-MF10}-\ref{fig-MF11} (c)),
  where their corresponding spectra (dotted points) are
  shown in Figs.~\ref{fig-MF10}-\ref{fig-MF11} (f).

  \section{Statistical Tests}
  \label{s-Random}

  There are several options available for analyzing the randomness of
  the pseudo random bit generators. The four most
  popular options according to Ref.~\refcite{Vinod} are the following: NIST suite of
  statistical tests,
  the DIEHARD suite of statistical
  tests, the Crypt-XS suite of statistical tests and the Donald
  Knuth's statistical test set.

  In this paper, we consider the NIST suite to
  analyze the generated pseudo random sequence keys.
  The main reason is that this suite has several appealing
  properties \cite{NIST,Charmaine}. For instance, it is uniform,
  it is composed by a number
  of well known tests and, for all of them, an exhaustive
  mathematical treatment is available. In addition, the source code
  of all tests in the suite is public available and is regularly
  updated \cite{NIST}.
  In fact, in Ref.~\refcite{NIST} is mentioned
  that the NIST
  suite may be useful as a first step in determining whether or not a
  generator is suitable for a particular cryptographic application.

  The NIST suite is a
  statistical package consisting of 15 tests that were developed to
  test the randomness of (arbitrarily long) binary sequences produced
  by either hardware or software based cryptographic random or
  pseudo-random number generators. These tests focus on a variety of
  different types of non-randomness that could exist in a sequence.
  Some tests are decomposable into a variety of subtests, and the 15 tests
  are listed in Table~\ref{tabla NIST}.

   \begin{table}[ht]
    \tbl{List of NIST Statistical Tests.}
    {
    \begin{tabular}{@{}cc@{}} \toprule
        Number & Test name \\
        \colrule
        1 & The Frequency (Monobit) Test \\
        2 & Frequency Test within a Block \\
        3 & The Runs Test \\
        4 & Tests for the Longest-Run-of-Ones in a Block \\
        5 & The Binary Matrix Rank Test \\
        6 & The Discrete Fourier Transform (Spectral) Test \\
        7 & The Non-overlapping Template Matching Test \\
        8 & The Overlapping Template Matching Test \\
        9 & Maurer's ``Universal Statistical'' Test \\
        10 & The Linear Complexity Test \\
        11 & The Serial Test \\
        12 & The Approximate Entropy Test \\
        13 & The Cumulative Sums (Cusums) Test \\
        14 & The Random Excursions Test \\
        15 & The Random Excursions Variant Test \\
        \botrule
    \end{tabular}
    \label{tabla NIST}
    }
    \end{table}

   For each statistical test, a set of $P-values$ (corresponding to the
   set of sequences) is produced. For a fixed significance level $\alpha$, a
   certain percentage of $P-values$ are expected to pass/fail the
   tests.
   For example, if the significance level is chosen to be 0.01 (i.e.,
   $\alpha = 0.01$), then about 1\% of the sequences are expected to
   fail. A sequence passes a statistical test whenever the $P-value
   \geq \alpha$  and fails otherwise. For each statistical test, the
   proportion of sequences that pass is computed and analyzed
   accordingly.
   It is not sufficient to look solely at the acceptance rates and
   declare that the generator be random if they seem fine. If the test
   sequences are truly random, the $P-values$ calculated are expected to
   appear uniform in $[0,1]$.

  For the interpretation of test results, NIST has adopted two
  approaches,  (1) the examination of the proportion of sequences that
  pass a statistical test and (2) the distribution of $P-values$ to
  check for uniformity.

   \begin{itemlist}
     \item  \emph{Proportions of the sequence passing the tests}:
           For each test is computed the proportion of sequences that passes the tests. First, the range of acceptable proportions is determined using the confidence interval, which is defined as

           \begin{equation}
               \label{eq-proportion}
               \hat{p} \pm 3 \sqrt { \frac{ \hat{p} (1 - \hat{p})} {m}},
     \end{equation}

      where $\hat{p} = 1-\alpha $, and $m$ is the sample size. If the proportion falls outside of this interval, then there is evidence that the data is non-random.

      \item \emph{Uniform Distribution of $P-values$}:
      The distribution of P-values is examined to ensure
      uniformity. This may be visually illustrated using a histogram.
      It may be also computed
      by means of a chi-square
      test ($\chi^2$), and the determination of a $P-value$
      of the $P-values$.
      The computation is as follows:

      \begin{equation}
                  \label{eq-distribution}
                  \chi^2=
                  \sum_{i=1}^{10}\frac{(f_i-\frac{m}{10})^2}{\frac{m}{10}},
        \end{equation}

      where $f_i$ is the number of $P-values$ in the sub-interval $i$ and
      $m$ is the size of the sample, which is $m=100$ for the present
      analysis. A $P-value$ is calculated such that
      $P-value_T={\rm igamc}(9/2,\chi^2/2)$,
      where ${\rm igamc}(n,x)$ is the incomplete gamma function. If
      $P-value_T \geq 0.0001 $,
      then the sequences can be considered to be uniformly
      distributed.

      \end{itemlist}

  \subsection{Results of the NIST statistical test suite}

    For the present statistical test, the two analyses  described
    above are applied and evaluated to determine if the
    generated sequences are random or not.
    We have considered $m=100$ samples of  $10^6$ bit sequences,
    where each sequence has been generated from a
    randomly chosen seed, and
    the proportion must lie above 0.960150($\alpha = 0.01$) and
    $P-value_T \geq 0.0001 $.
    In order to investigate the performance of the generator,
    we analyze the generated pseudo-random sequences for
    $N=7$, $N=15$ and $N=31$ bits,
    considering one and three transformations.

    \subsubsection{Case of $N=7$ bits}

\begin{figure}[ph]
      \centerline{\psfig{file=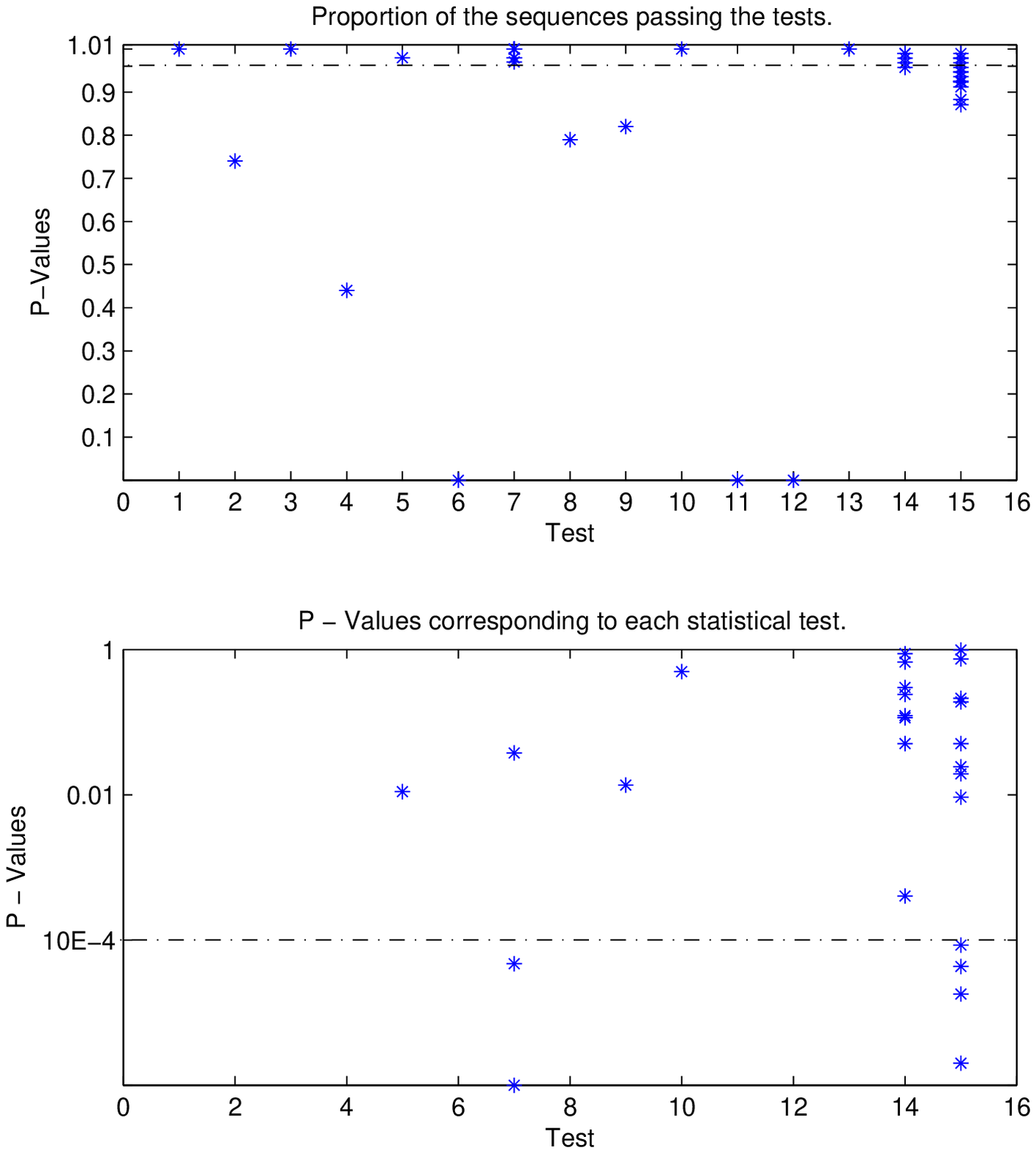,width=10cm, height=10cm}}
      \vspace*{8pt}
      \caption{Proportions and $P-values_T$ corresponding to $N=7$
               bits and one transformation. Dashed line
               separates the success and failure regions.}
         \label{fig-1R7bits}
       \end{figure}

\begin{figure}[ph]
      \centerline{\psfig{file=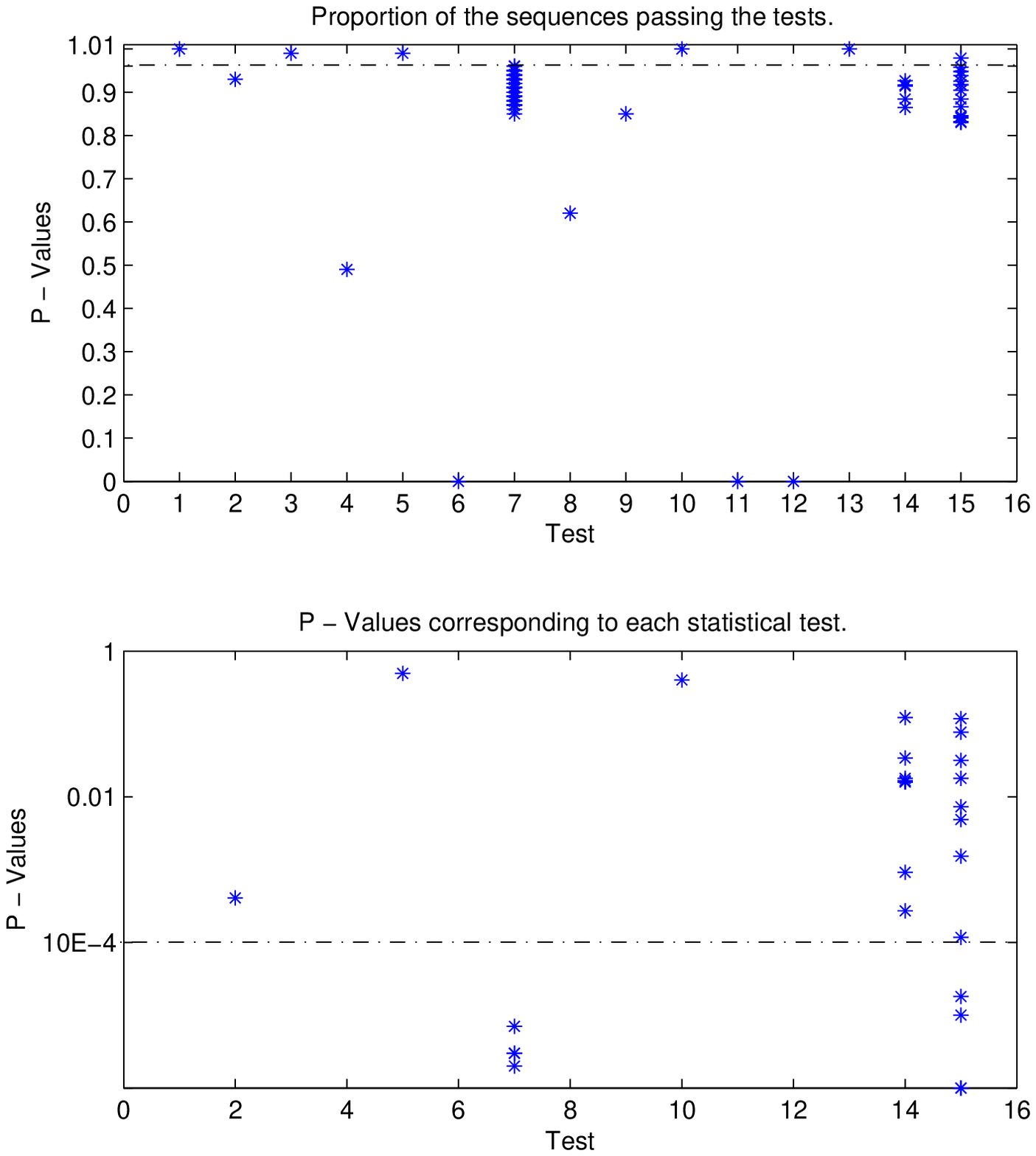,width=10cm, height=10cm}}
      \vspace*{8pt}
      \caption{Proportions and $P-values_T$ corresponding to $N=7$
               bits and three transformations. Dashed line
               separates the success and failure regions.}
         \label{fig-3R7bits}
       \end{figure}

    In Figs.~\ref{fig-1R7bits}-\ref{fig-3R7bits}
    are shown the results from the NIST testing
    for $N=7$ bits for one and three transformations, respectively.
    We can observe that in this case
    there is a poor performance;
    the generated pseudo random sequences just passes
    some tests and are not uniformly distributed.

  \subsubsection{Case of $N=15$ bits}

\begin{figure}[ph]
      \centerline{\psfig{file=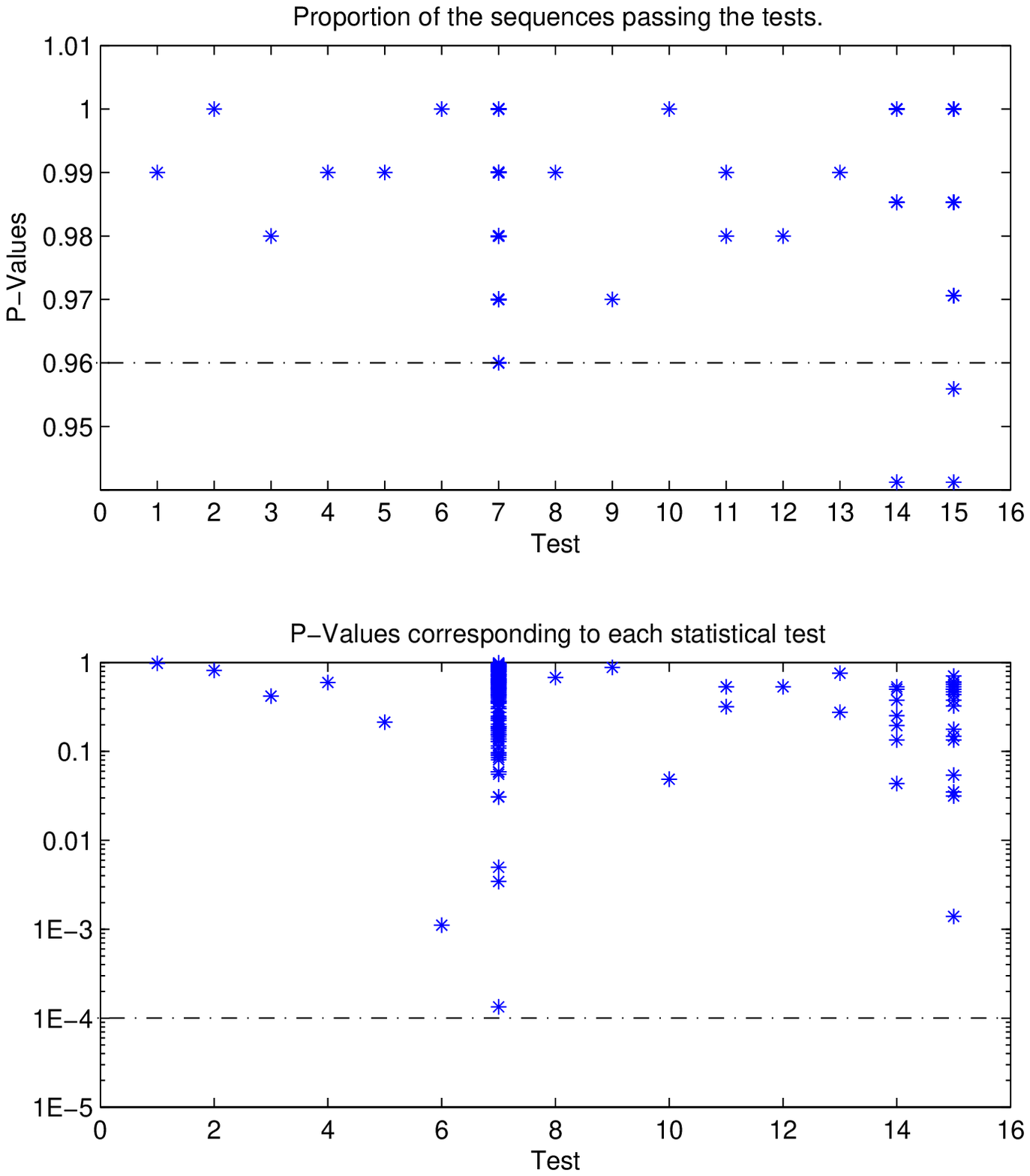,width=10cm, height=10cm}}
      \vspace*{8pt}
      \caption{Proportions and $P-values_T$ corresponding to $N=15$
               bits and one transformation. Dashed line
               separates the success and failure regions.}
         \label{fig-1R15bits}
       \end{figure}

\begin{figure}[ph]
      \centerline{\psfig{file=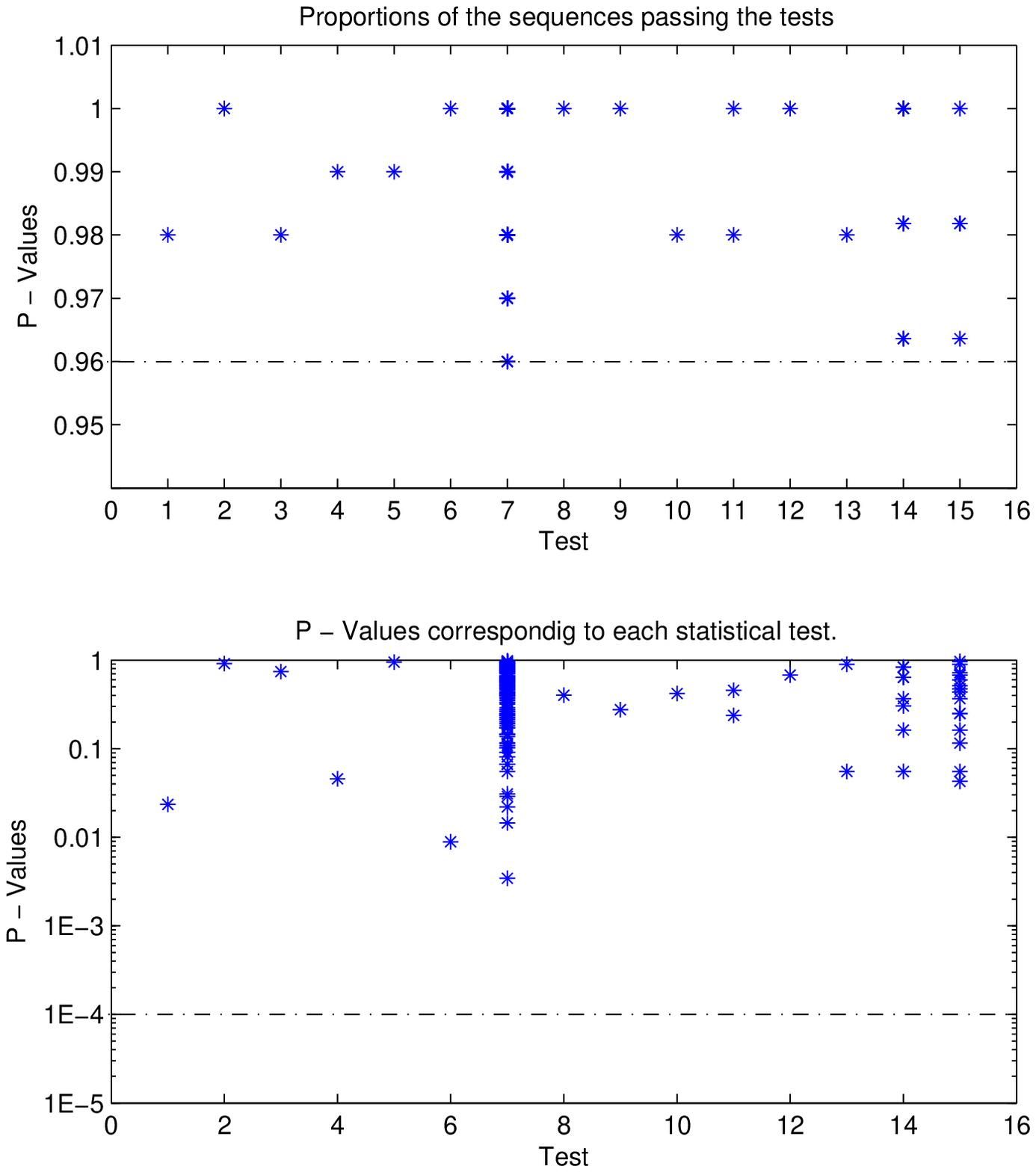,width=10cm, height=10cm}}
      \vspace*{8pt}
      \caption{Proportions and $P-values_T$ corresponding to $N=15$
               bits and three transformations. Dashed line
               separates the success and failure regions.}
         \label{fig-3R15bits}
       \end{figure}

    Figs.~\ref{fig-1R15bits}-\ref{fig-3R15bits}
    show the results for $N=15$ bits for one
    and three transformations, respectively.
    There is better performance that in the previous
    case. We can observe that using one transformation,
    the generator do not passes all tests, see
    Fig.~\ref{fig-1R15bits}, but it is
    uniformly distributed.

    \subsubsection{Case of $N=31$ bits}

\begin{figure}[ph]
      \centerline{\psfig{file=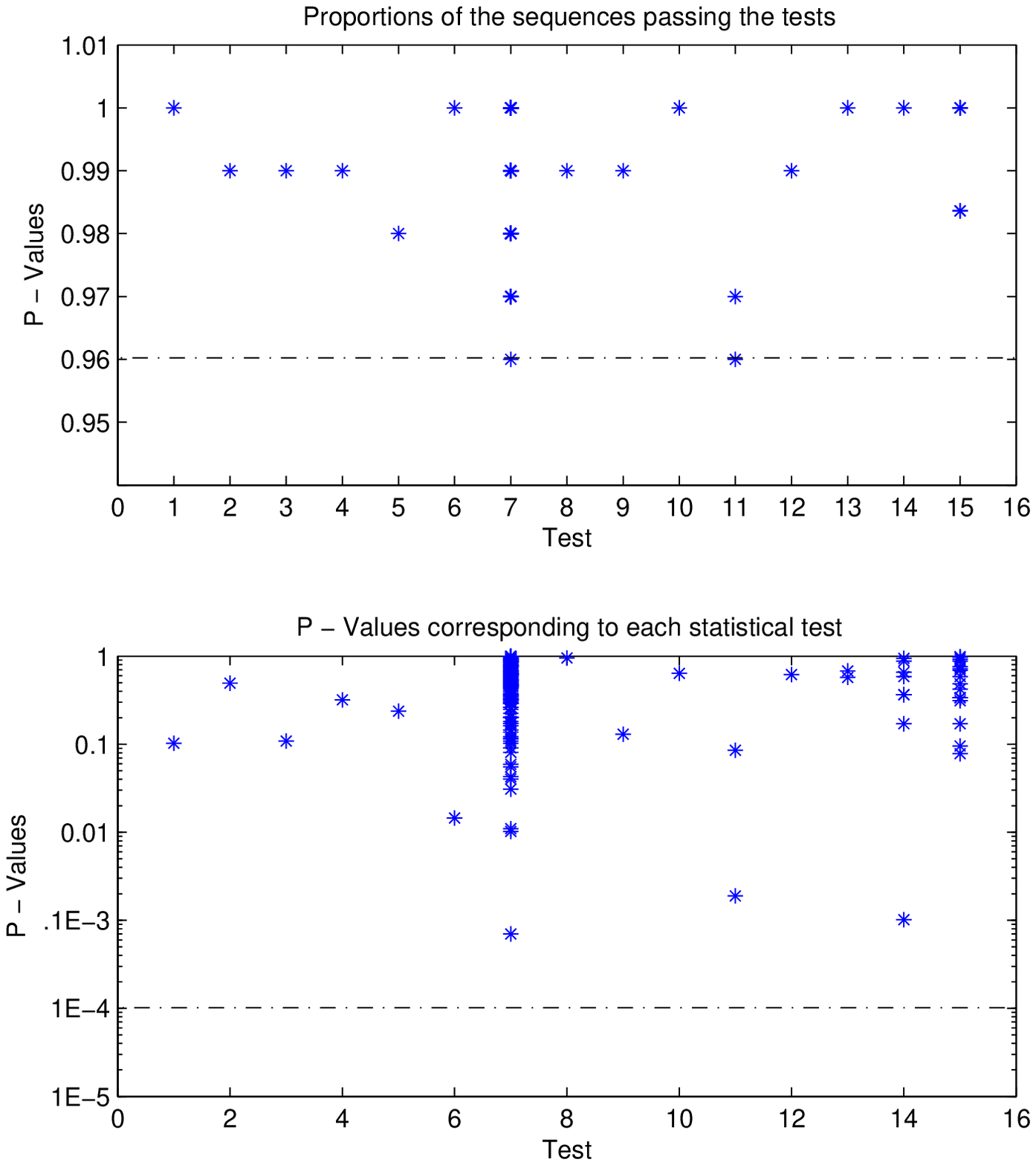,width=10cm, height=10cm}}
      \vspace*{8pt}
      \caption{Proportions and $P-values_T$ corresponding to $N=31$
               bits and one transformation. Dashed line
               separates the success and failure regions.}
         \label{fig-1R31bits}
       \end{figure}

\begin{figure}[ph]
      \centerline{\psfig{file=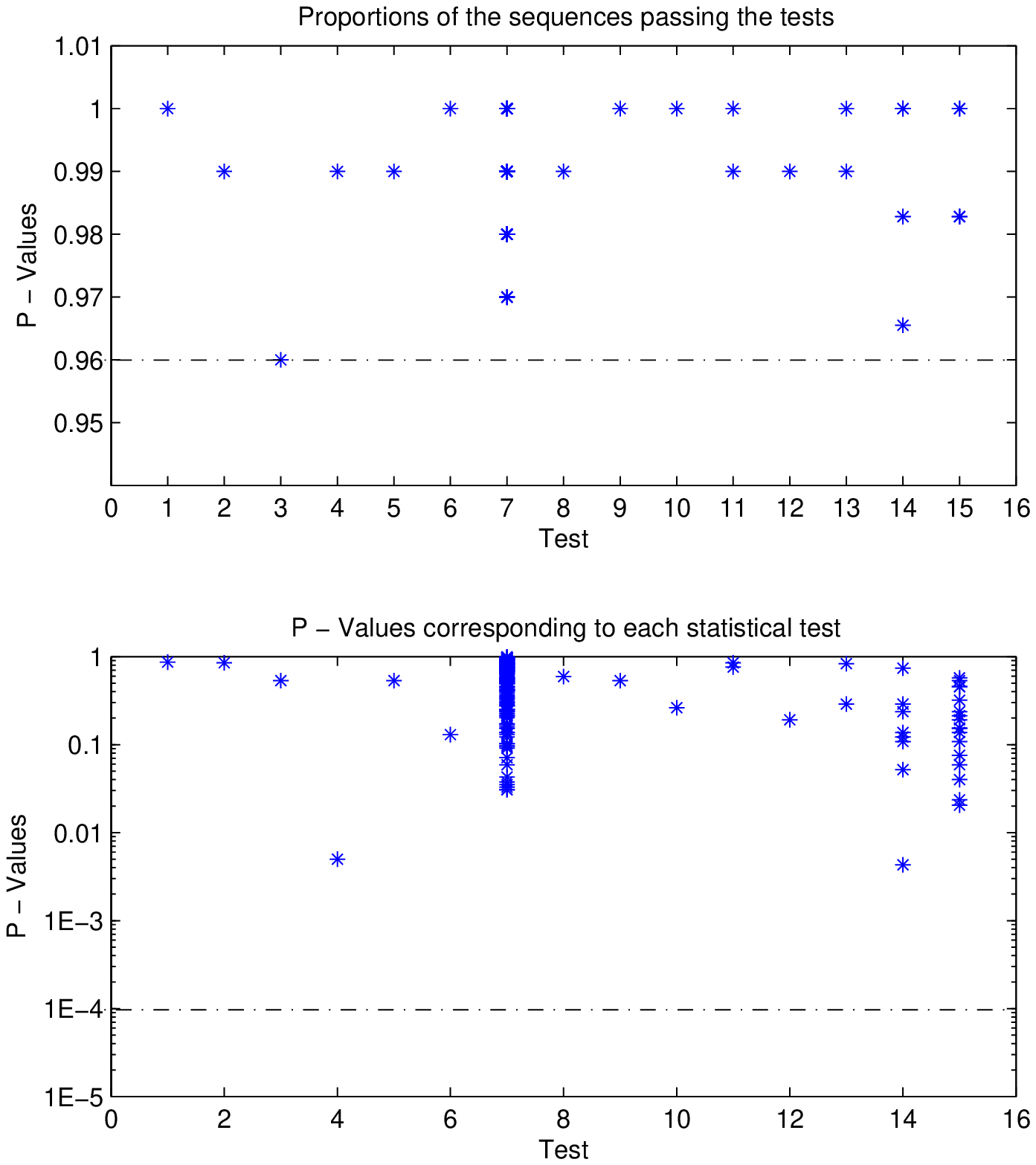,width=10cm, height=10cm}}
      \vspace*{8pt}
      \caption{Proportions and $P-values_T$ corresponding to $N=31$
               bits and three transformations. Dashed line
               separates the success and failure regions.}
         \label{fig-3R31bits}
       \end{figure}

    In the last case, Figs.~\ref{fig-1R31bits}-\ref{fig-3R31bits}
    show the results for $N=31$ bits for one
    and three transformations, respectively.
    As we can see, all tests are passed using
    one and three transformations.

    \section{Conclusions}

      We have implemented and reviewed a pseudo-random number
      generator based on a rule-90 cellular automata. This generator
      in its basic form (using one transformation), and its
      modified version (with three transformations) are
      analyzed by means of a sequence matrix $H_N$.
      The intrinsic multifractal properties of the sequence matrix in the two versions of the generator
      are discussed having in mind their possible usage in cryptanalysis. In addition, the quality of the generated pseudo random sequences
      are evaluated using the NIST statistical tests. According to these tests, this PRNG can generate
      high-quality random numbers using one or three transformations.
      It is worth noticing that the longer the length of the generated numbers the better is the quality of the random numbers we obtain.
      In other words, the generator will produce sequences of keys with a better quality as the size
      of keys is increased.
      The only case that fails to pass all the tests with one or three transformations is
      for $N=7$ bits. There are also some statistical problems for $N=15$ bits using one transformation.
      We were able to obtain random sequences of 15 bits, without
      repeating, of period length $2^{27}$ and $2^{31}$ using one and three transformations, respectively.
      These results are sufficient for many cryptographic applications.

    \section*{Acknowledgments}
      G. Flores-Era\~na is a doctoral fellow of CONACYT (M\'exico) in
      the program of ``Ciencias Aplicadas'' at IICO-UASLP.

\end{document}